\documentstyle[12pt,epsf]{article}

\newcommand{\eps}{\varepsilon}
\newcommand{\sla}{\not\;\!\!\!}
\newcommand{\skal}{\!\cdot\!}

\newcommand{\lvec}[1]{\stackrel{\bf \leftarrow}{#1}}

\begin{document}
{\Large \bf QCD sum rules with finite masses}

\begin{quote}
\small
{\bf M.\,Meyer-Hermann, A.\,Sch\"afer and W.\,Greiner}\\[2mm]
{\it Institut f\"ur Theoretische Physik der Johann Wolfgang
 Goethe-Universit\"at, D-60054 Frankfurt, Germany}\\[2mm]
The concept of QCD sum rules is extended to bound states
composed of particles with finite mass such as scalar quarks or
strange quarks. It turns out that mass corrections become
important in this context. The number of relevant corrections is
analyzed in a systematic discussion of the IR- and UV-divergencies,
leading in general to a finite number of corrections. The results are
demonstrated for a system of two massless quarks and two heavy scalar
quarks.
\end{quote}

\section{Introduction}

In the last fifteen years the concept of QCD sum rules invented by
Shifman, Vain\-shtein and Zakharov \cite{shifman}
has been very well developed.
The intention was to
include non-perturbative effects in QCD calculations, which naturally
become important when dealing with hadron physics.
The concept of vacuum condensates is best illustrated in the case of
chiral symmetry, which is spontaneously broken for hadrons
as one can see by comparing the masses of different mesons with
identical quantum numbers (i.e. $\rho$ and $a_1$ meson). A direct
consequence of spontaneous breakdown of any symmetry is the appearance
of non-vanishing vacuum condensates. The
idea of QCD sum rules is to parametrize the non-trivial QCD vacuum with
vacuum condensates, including the chiral condensate $<\overline{q}q>$,
and in this way to realize a parametrization of
non-perturbative effects in QCD.

The condensates appear as matrix elements in the
operator-product-expansion \cite{wilson} (OPE) of a physical function
(i.e. the polarization function):
\begin{eqnarray} \label{kap1n1}
\Pi(q^2) \sim \sum^\infty_{j=1} <0|B_j(0)|0> \int d^4x\;E_j(x)\,e^{iqx}
\end{eqnarray}
This expansion separates perturbative (Wilson coefficents
$E_j(x)$) and non-per\-tur\-ba\-tive (condensates $<0|B_j(0)|0>$) parts, being
thus
completely calculable with perturbative Feynman rules.
It can be related to the hadronic spectrum through a
dispersion relation with one subtraction $C$
\begin{equation} \label{kap1n2}
\Pi_{OPE}(q^2)
= \frac{q^2}{\pi}
  \int^\infty_0 ds\;\frac{Im\,\Pi_{Had}(s)}{s(s-q^2)} + C
\end{equation}
where the imaginary part of the function $\Pi_{Had}$ is connected to
the hadronic spectrum.
In that way it is possible to get
hadronic quantities like the mass of a bound state as a function
of a few vacuum condensates, which are fitted to a large number of
hadronic data.

The kind of condensates being relevant in the OPE depends strongly on the
flavour composition and the quantum numbers of the bound state
considered. In the case of light
quarks (up or down) there is strong evidence for the presence of
quark-antiquark pairs
in the vacuum, so that apart from the gluon-condensates the
quark-condensates have to be taken into account. If the
particles are heavy quarks (charm or heavier) the value of the
corresponding condensates becomes very small and may be neglected, so
that the calculation may lead to a reasonable result by including the
gluon-condensate only.
The strange quark, however, poses a problem because it is neither
light nor
heavy. Therefore its condensate can not be neglected a priori, and in
order to take it into account, it is necessary to treat massive
quarks in QCD sum rules --- a problem unsolved up to now. We derive
several results necessary to approach this goal and elucidate the problems
occuring.

This article is organized as follows.
We shall illustrate the problem of finite mass sum rules with
a specific example (introduced in section $2$),
a detailed study of which will be published elsewhere.
In the third section we give an
expression for the matrix elements of normal-ordered, nonlocal field
products --- which occurs in the expansion of the correlator under
consideration ---, demonstrating that the finite masses
cause corrections to the commonly used expressions. Afterwards some of
the Wilson coefficients introduced above will be discussed with
special attention to the appearing infrared and ultraviolett
divergencies and to the relevance for the sum rule.


\section{The system under consideration}
For studying the influence of mass terms in
QCD sum rules calculations we will consider a four-particle
bound state composed of two light fermionic quarks and two heavy
scalar quarks. The last ones are hypothetical particles introduced in
QCD phenomenologically. Scalar quarks are predicted by the
GUT-theory super\-sym\-me\-try \cite{sohnius} (they are called squarks),
such that we are not merely dealing with a toy model. The physical problem
motivating this model is wether or not the $2$ quark -- $2$ squark
bound state may be energetically more favourable than a $2$ squark
state, which is predicted to have a very high mass \cite{baer...}.
The lowest lying $2$ quark -- $2$ squark bound state can be
calculated --- using QCD sum rules --- in dependence of different new
condensates and of the squark mass. These parameters are unknown, but
if one could find a reasonable choice of them, such that the
$2$ quark -- $2$ squark bound state mass becomes small, this would be
most interesting for supersymmetry phenomenology. The results of
these calculations will be published elswhere.

\begin{figure}[htb]
\center{\hspace*{0mm} \epsfxsize=9.3cm \epsfysize4cm
        \epsfbox{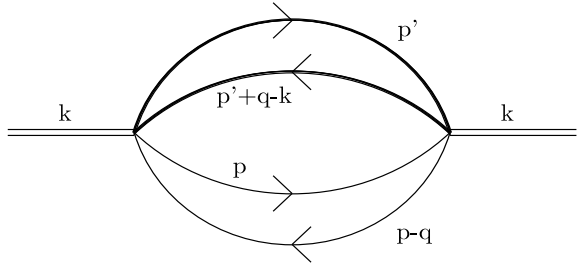}}
\caption{\label{feyn-pert} Perturbative Feynman diagram with two scalar
(normal lines) and two fermion propagators (heavy lines)}
\end{figure}

In the following calculation the quark masses are neglected
and the scalar quarks are treated as massive particles.
In order to calculate the mass of the lowest lying bound state of
this system, the OPE of the
polarization-function corresponding to the diagram in
Figure \ref{feyn-pert} has to be considered (see eq.\,(\ref{kap1n1})),
which is given by the two-point-function
\begin{eqnarray} \label{pol13}
\Pi(k^2)
\;=\; -\frac{ig^2}{3} \int d^4(x-y)\;e^{ik(x-y)}\,
      g_{\mu\nu} <0| T\{ J^\mu(x),J^\nu(y) \}|0>
\end{eqnarray}
where $T$ denotes the time ordered product.
The current $J^\mu(x)$ contains an incoming and outgoing
fermionic and scalar quark:
\begin{eqnarray} \label{pol4}
J^\mu (x) = g\;\overline{\psi}(x)\gamma^\mu \psi(x)\;
               \overline{\phi}(x) \phi(x) \quad.
\end{eqnarray}
The present sum rule is calculated in lowest order of QCD perturbation
theory. Thus the exchange of gluons and the coupling to the
gluon condensate are neglected.

Usually QCD sum rule calculations are carried out in configuration
space. In the case of massless particles this choice simplifies
the calculations: the propagators are as simple as in momentum space,
but the number of integrations is reduced to one, while in momentum
space the number of integrations is equal to the number of loops
in the corresponding Feynman diagram. However, in the case of
particles with non-vanishing masses the propagators become very
cumbersome, so that in most cases calculations in
momentum space will be more convenient.

The next step is to find the explicit form of the OPE for the
polarization function (\ref{pol13}). The time-ordered product
has to be expanded into normal-ordered products with attention to
the non-vanishing vacuum condensates. In this way one gets sixteen
terms, each of them corresponding to one diagram: first there is the
perturbative diagram (see figure \ref{feyn-pert}),
which corresponds to the fully contracted term without any condensates
\begin{eqnarray} \label{term1}
-\;<0\mid  Tr\{S(x-y) \gamma^\mu S(y-x) \gamma^\nu\}\;
           Tr\{\Delta(y-x) \Delta(x-y) \}\mid0>
\end{eqnarray}
where $S(x-y)$ is the Feynman-propagator of the quarks
and $\Delta(x-y)$ is the scalar quark propagator.
The traces are taken over color- and
Dirac-indices in the case of fermionic propagators and in color space
only in the case of scalar propagators.

There are eight terms containing one non-contracted pair of fermionic
field operators of the form
\begin{eqnarray}
+\,i <0\mid \left(\gamma^\nu\right)_{ik} S_{kl}(y-x)
            \left(\gamma^\mu\right)_{lj} \,
           :\overline{\psi}_{\alpha i}(y) \psi_{\beta j}(x):\,
            Tr\{\Delta(y-x) \Delta(x-y) \}\mid0>
\end{eqnarray}
which all vanish independently of the scalar part of the term:
the condensate of massless fermions is proportional to $\delta_{ij}$,
such that a trace over an odd number of $\gamma$-matrices appears in
the fermionic part of the expression.

The term without any contraction is vanishing as well, because it
contains no propagator, so that there is no momentum flow
through the diagram.

Six diagrams are left with one or two non-contracted pairs
of scalar field operators, for example the contribution from figure
\ref{scalarkonden}
\begin{eqnarray} \label{term3}
-\;i<0\mid  Tr\{S(x-y) \gamma^\mu S(y-x) \gamma^\nu\}\;
            \Delta(x-y) \;
           :\overline{\phi}(x) \phi(y): \mid0> \quad .
\end{eqnarray}
When expanding expressions like that given in eq.\,(\ref{pol13}) one
quite generally encounters
{\it nonlocal} normal ordered product of scalar or fermionic
field operators.
In these products the field operators are taken at
different points in Minkowski space, while the vacuum condensates of
scalar or fermionic fields are
defined at the same points. Nevertheless, the nonlocal
normal ordered products commonly are identified with the
vacuum condensates. We claim that this procedure is correct only
in zeroth order in the mass of the particles under consideration.


\section{Mass corrections to the condensates}
\subsection{Condensates of scalars}
\label{scalar}
In the previous section we have shown, that vacuum expectation values
of normal ordered products of scalar fields at different space-time
points of the type
\begin{eqnarray} \label{cond1}
 <0\mid :\overline{\phi}_\alpha (x) \phi_\beta (y): \mid0>
\end{eqnarray}
occur in the operator product expansion. Here the greek letters denote
the color space indices important in the following. With the object
of establishing a relation between the expression (\ref{cond1})
and the vacuum condensates defined by
\begin{eqnarray} \label{cond2}
<\overline{\phi}\phi>\;
:=\;\sum_{\alpha\beta}
  <0|:\overline{\phi}_{\alpha}(x)\phi_{\beta}(x):|0>
\end{eqnarray}
we will have to expand one of the fields in eq.\,(\ref{cond1}) with
respect to the
space-time point $y$ belonging to the second field. The corresponding
Taylor series is
\begin{eqnarray} \label{cond3}
\overline{\phi}_\alpha(x)
\;=\;\sum_{n=0}^\infty \frac{1}{n!}(x-y)^{\mu_1}\cdots(x-y)^{\mu_n}
     \Big(\overline{\phi} \lvec{\partial}_{\mu_1}\cdots
                          \lvec{\partial}_{\mu_n}
     \Big)_\alpha (y) \quad .
\end{eqnarray}
To preserve gauge invariance the ordinary derivatives must be
replaced by the covariant ones.
\begin{eqnarray} \label{cond4}
\overline{\phi}_\alpha(x)
\;=\;\sum_{n=0}^\infty \frac{1}{n!}(x-y)^{\mu_1}\cdots(x-y)^{\mu_n}
     \Big(\overline{\phi} \lvec{D}_{\mu_1} \cdots \lvec{D}_{\mu_n}
     \Big)_\alpha (y)
\end{eqnarray}
The replacement is allowed if there exists a gauge for which
the above two expressions are equivalent. The
coordinate gauge $x^\mu A^a_\mu = 0$ has exactly this property (see
for example \cite{pascual}).

Using this expansion we will derive an expression of the full series
describing the nonlocal field product in eq.\,(\ref{cond1}):
\begin{eqnarray} \label{cond5}
&&\hspace*{-1cm}<0|:\overline{\phi}_\alpha(x)\phi_\beta(y):|0>
  \nonumber\\
&=& <0|:\overline{\phi}_\alpha(y)\phi_\beta(y):|0> \nonumber\\
&&+ <0|:\left(\overline{\phi}\lvec{D}_{\mu_1}\right)_\alpha(y)
        \phi_\beta(y):|0>\;\xi^{\mu_1} \nonumber\\
&&+ \,\frac{1}{2}<0|:\left(\overline{\phi}
                           \lvec{D}_{\mu_1}\lvec{D}_{\mu_2}\right)
                        _\alpha(y)
        \phi_\beta(y):|0>\;\xi^{\mu_1}\xi^{\mu_2} \nonumber\\
&&+\,\cdots\nonumber\\
&&+ \,\frac{1}{m!}<0|:\left(\overline{\phi}
                            \lvec{D}_{\mu_1}\cdots \lvec{D}_{\mu_m}\right)
                        _\alpha(y)
        \phi_\beta(y):|0>\;\xi^{\mu_1}\cdots\xi^{\mu_m}\nonumber\\
&&+\,\cdots
\end{eqnarray}
where $\xi:=x-y$.

The terms in eq.\,(\ref{cond5}) can contribute to the series only if
the corresponding operators do not have any uncontracted Lorentz index,
for the vacuum state is a scalar state.
Consequently only the terms with an even number of
covariant derivatives are of interest (the remaining
terms will be renumbered using $n:=m/2$).
All color and Lorentz indices must be contracted in
this terms, so that the structure of the vacuum expectation value in
color and Lorentz space is fixed a priori:
\begin{eqnarray} \label{cond6}
<0|:\overline{\phi}_\alpha(y)\phi_\beta(y):|0>
&\!=&\! C_0 \,\delta_{\alpha\beta} \nonumber\\
<0|:\left(\overline{\phi}\!\lvec{D}_{\mu_1}\lvec{D}_{\mu_2}
    \right)_{\!\alpha}\!\!(y)
        \phi_\beta(y):|0>
&\!=&\! C_1 \,g_{\mu_1\mu_2}\delta_{\alpha\beta} \nonumber\\
<0|\!\!:\!
   \left(\overline{\phi}\!\lvec{D}_{\mu_1}\cdots
                        \lvec{D}_{\mu_4}\right)_{\!\alpha}\!\!(y)
        \phi_\beta(y)\!:\!\!|0>
&\!=&\! C_2 \big(g_{\mu_1\mu_2}g_{\mu_3\mu_4}
            \!+\!g_{\mu_1\mu_3}g_{\mu_2\mu_4}
            \!+\!g_{\mu_1\mu_4}g_{\mu_2\mu_3}
            \big)\delta_{\alpha\beta} \nonumber\\
<0|\!\!:\!
   \left(\overline{\phi}\!\lvec{D}_{\mu_1}\cdots
                        \lvec{D}_{\mu_{2n}}\right)_{\!\alpha}\!\!(y)
        \phi_\beta(y)\!:\!\!|0>
&\!=&\! C_n \underbrace{
           \big(g_{\mu_1\mu_2}g_{\mu_3\mu_4}\cdots g_{\mu_{2n-1}\mu_{2n}}
           +\mbox{perm.}
           \big)}_{(2n-1)!!\quad\mbox{terms}}
    \delta_{\alpha\beta} \nonumber\\
\end{eqnarray}
The coefficients $C_n$ are to be calculated by contracting the
equations (\ref{cond6}) with the metric tensor $g^{\mu_1\mu_2}$,
using the Klein-Gordon field equation
$(\Box+m^2)\phi(y)\;=\;0$ with
$\Box:=g^{\mu\nu}D_\mu D_\nu$, and summing over all remaining
indices. By this procedure one gets the condensates (\ref{cond2})
multiplied with the mass in the corresponding dimension on
the left hand side and the coefficient of interest with a
numerical factor on the right hand side. For instance:
\begin{eqnarray} \label{cond7}
<\overline{\phi}\phi>
&=& \sum_{\alpha\beta}
    <0|:\overline{\phi}_\alpha(y)\phi_\beta(y):|0> \nonumber\\
&=& C_0\sum_{\alpha\beta}\delta_{\alpha\beta}
\;=\; N_c C_0\nonumber\\
\Longrightarrow\quad C_0 &=& \frac{1}{N_c}<\overline{\phi}\phi> \\
\label{cond8}
-m^2<\overline{\phi}\phi>
&=& \sum_{\alpha\beta}
    <0|:\left(\overline{\phi}
          g^{\mu_1\mu_2}\lvec{D}_{\mu_1}\lvec{D}_{\mu_2}
        \right)_\alpha(y)
        \phi_\beta(y):|0>\nonumber\\
&=& C_1\sum_{\alpha\beta}\delta_{\alpha\beta}
       g^{\mu_1\mu_2}g_{\mu_1\mu_2}
\;=\; 4N_c C_1\nonumber\\
\Longrightarrow\quad C_1 &=& -\frac{m^2}{4N_c}<\overline{\phi}\phi> \\
\label{cond9}
m^4<\overline{\phi}\phi>
&=& \sum_{\alpha\beta}
    <0|:\left(\overline{\phi}
          g^{\mu_1\mu_2}g^{\mu_3\mu_4}
          \lvec{D}_{\mu_1} \lvec{D}_{\mu_2} \lvec{D}_{\mu_3} \lvec{D}_{\mu_4}
        \right)_\alpha(y)
        \phi_\beta(y):|0>\nonumber\\
&=& C_2\sum_{\alpha\beta}\delta_{\alpha\beta}
       g^{\mu_1\mu_2}g^{\mu_3\mu_4}
       \left(g_{\mu_1\mu_2}g_{\mu_3\mu_4}
            +g_{\mu_1\mu_3}g_{\mu_2\mu_4}
            +g_{\mu_1\mu_4}g_{\mu_2\mu_3}
       \right)\nonumber\\
&=& N_c C_2 \left(16 + g^{\mu_2}_{\;\mu_3}g^{\;\mu_3}_{\mu_2}
                     + g^{\mu_2}_{\;\mu_4}g^{\;\mu_4}_{\mu_2}  \right)
\;=\; 24N_c C_2\nonumber\\
\Longrightarrow\quad C_2 &=& \frac{m^4}{24N_c}<\overline{\phi}\phi>
\end{eqnarray}
where $N_c$ is the number of colors.

The general form of $C_{n>2}$ is
\begin{eqnarray} \label{cond10}
C_n &=& (-1)^n \frac{m^{2n}}{N_c G_n}<\overline{\phi}\phi> \quad,
\end{eqnarray}
with
\begin{eqnarray} \label{cond11}
G_n:=g^{\mu_1\mu_2}\cdots g^{\mu_{2n-1}\mu_{2n}}
     \left(g_{\mu_1\mu_2}\cdots g_{\mu_{2n-1}\mu_{2n}}
           + \mbox{permutations}\right) \quad .
\end{eqnarray}
For each contraction of two covariant derivatives one gets a factor
$-m^2$ coming from the Klein-Gordon field equation, all terms contain the
same factor $N_c$ from the summation in color space, and the factor
$G_n$ is appearing by construction. One can easily verify that the
relation $G_n=2^n(n+1)!$ holds for all $n$.

In this way the coefficients $C_n$ of eq.\,(\ref{cond10}) are
known explicitly and thus we have found an expression for all the vacuum
expectation values of eq.\,(\ref{cond6}). Inserting this into
the expansion (\ref{cond5}) we get:
\begin{eqnarray} \label{cond12}
<0|:\overline{\phi}_\alpha(x)\phi_\beta(y):|0>
&=& \sum_{n=0}^\infty (-1)^n
\frac{1}{(2n)!}
\frac{m^{2n}}{2^n N_c (n+1)!}
<\overline{\phi}\phi> \delta_{\alpha\beta} \nonumber\\
&& \times
\underbrace{\big(g_{\mu_1\mu_2}g_{\mu_3\mu_4}\cdots g_{\mu_{2n-1}\mu_{2n}}
    +\mbox{perm.} \big)}_{(2n-1)!!\quad\mbox{terms}}
\xi^{\mu_1}\cdots\xi^{\mu_{2n}} \nonumber\\
\end{eqnarray}
The $(2n-1)!!$ terms contract the $2n$ dimensional tensor
$\xi^{\mu_1}\cdots\xi^{\mu_{2n}}$ yielding $(2n-1)!!(\xi^2)^n$, so that we
are left with the full expression of the {\it nonlocal} field product
in eq.\,(\ref{cond1}) to all orders in the mass:
\begin{equation} \label{cond13}
<0|:\overline{\phi}_\alpha(x)\phi_\beta(y):|0>
\;=\; \frac{1}{N_c}\delta_{\alpha\beta}<\overline{\phi}\phi>
\sum_{n=0}^\infty (-1)^n \frac{1}{n!(n+1)!}
\left(\frac{m^2\xi^2}{4}\right)^n
\quad,
\end{equation}

The sum can be performed giving:
\begin{equation} \label{cond15}
<0|:\overline{\phi}_\alpha(x)\phi_\beta(y):|0>
\;=\; \frac{1}{N_c}\delta_{\alpha\beta}<\overline{\phi}\phi>
\frac{2}{m|\xi|}\,J_1\left(m|\xi|\right)
\end{equation}
This result is not so surprising, because there is a connection
between the scalar propagator and the Bessel functions. The left hand
side of eq.\,(\ref{cond15}) has the structure of a scalar
propagator as well. In this sense it may be meaningful to call the
vacuum expectation value of a nonlocal normal ordered product of two
fields a {\it non-perturbative propagator}.

In the case of vanishing mass the series reduces to the commonly used
relation
\begin{eqnarray} \label{cond14}
<0|:\overline{\phi}_\alpha(x)\phi_\beta(y):|0>
\;=\; \frac{1}{N_c}\delta_{\alpha\beta}<\overline{\phi}\phi>
\end{eqnarray}
For nonvanishing mass the higher-order terms of the series have to be
taken into account. The question arises wether the series is
convergent and which order still is important in a specific calculation?
The answer depends strongly on the value $m^2 \xi^2$ where $\xi^2$
is the distance between the two vertices at $x$ and $y$, which is
inversely proportional to the mass $M^2$ of the incoming state to be
evaluated by QCD sum rules. So the ratio $m/M$ of the
composite particle mass and the bound state mass determine the importance of
the higher-order terms. For instance $m/M=1$ leads to a $10\%$ correction
in first order, while $m/M=2$ blows up the first order correction to
$50\%$ and the second order term makes another $10\%$ contribution.

To conclude: the
higher-order corrections may be neglected for systems where a heavy
bound state is constructed from light particles. However, the
higher-order corrections become dominant
if the bound state is lighter than its constituent particles.
For instance this could happen in the supersymmetric system introduced
in the second section. In the last case the exact form (\ref{cond13})
or (\ref{cond15}) has to be used.

\subsection{Condensates of fermions}
A treatment analogous to eq.\,(\ref{cond1}) is possible also for
fermion fields instead of the scalar fields used in the previous section.
This is especially important if the interest lies in the treatment of
finite masses in QCD sum rules emerging in mass calculations of
strange baryons \cite{belyaev}.

The following nonlocal normal-ordered product of fermion field
operators is to be expanded
\begin{eqnarray} \label{ferm1}
 <0\mid :\overline{\psi}_{i\alpha} (x) \psi_{j\beta} (y): \mid0>
\quad .
\end{eqnarray}
Here the fields carry a Dirac index (latin letter) in addition to
the color index (greek letter). The fermion condensate is defined by
\begin{eqnarray} \label{ferm2}
<\overline{\psi}\psi>\;
:=\;\sum_{ij\alpha\beta}
  <0|:\overline{\psi}_{i\alpha}(x)\psi_{j\beta}(x):|0>
\end{eqnarray}
and the corresponding expansion of one fermion field preserving gauge
invariance reads (see eq.\,(\ref{cond4}))
\begin{eqnarray} \label{ferm3}
\psi_{i\alpha}(y)
\;=\;\sum_{n=0}^\infty \frac{1}{n!}(y-x)^{\mu_1}\cdots(y-x)^{\mu_n}
     \Big(D_{\mu_1} \cdots D_{\mu_n} \psi
     \Big)_{i\alpha} (x)
\end{eqnarray}

This series can be evaluated in close analogy to the case of the
scalar field treated in sect.\,\ref{scalar}. There is one important
difference:
In the scalar case only the terms in the expansion of the field with
an even number of derivatives give a contribution to the series (see
discussion after eq.\,(\ref{cond5})),
while in the case of fermion fields the terms with an odd number of
derivatives are non-vanishing, for the Dirac equation
$(i\sla\! D - m)\psi = 0$ is of first order in the derivatives.

The Lorentz structure of the even terms is the one shown in
eq.\,(\ref{cond6}) with an additional factor $\delta_{ij}$. Each odd term
has the same structure as the corresponding even term with one more
derivative, except for the replacement of one $g_{\mu\nu}$
by $\gamma_\mu$. It turns out that the
coefficients $C_n$ ($n\in {0,1,2,\ldots}$) corresponding to
eq.\,(\ref{cond10}) are
\begin{eqnarray} \label{ferm4}
C_n &=& \frac{(-im)^n}{4N_c G_n}<\overline{\psi}\psi>
\end{eqnarray}
where the even coefficients $G_{2n}$ coincide with those
for scalar fields while the odd coefficients are given by
$G_{2n-1}=G_{2n}$. This means that the coefficient of a term
with an odd number of derivatives is equal to the coefficient of the
following term with an even number of derivatives (except a factor
$im$), which are determined by $G_{2n}=2^n(n+1)!$ like in the case of scalar
fields. After some algebra the whole series corresponding to
eq.\,(\ref{cond13}) reads:
\begin{equation} \label{ferm5}
\begin{array}{rcl}
&&
<0|:\overline{\psi}_{i\alpha}(x)\psi_{j\beta}(y):|0> \nonumber\\
&&
=\;\frac{\delta_{\alpha\beta}\delta_{ij}}{4N_c}<\overline{\psi}\psi>
\sum_{n=0}^\infty (-1)^n \frac{1}{n!(n+1)!}
                  \left(\frac{m^2\xi^2}{4}\right)^n \nonumber\\
&&
+\;\frac{im}{2}\,\frac{\delta_{\alpha\beta}\delta_{ij}}{4N_c}
               <\overline{\psi}\psi>\, \sla\xi\;
\sum_{n=1}^\infty (-1)^n \frac{1}{(n-1)!(n+1)!}
                  \left(\frac{m^2\xi^2}{4}\right)^{n-1}
\end{array}
\end{equation}
This result can be verified by acting with $(+i\sla\partial+m)$ on
the series (\ref{cond13}) for scalar fields
\begin{eqnarray} \label{ferm6}
 <0\mid :\overline{\psi}_{i\alpha} (x) \psi_{j\beta} (y): \mid0>
 \;\longrightarrow\; (i\sla\partial+m)
 <0\mid :\overline{\phi}_\alpha (x) \phi_\beta (y): \mid0>
\end{eqnarray}
with the replacement
\begin{equation} \label{ferm7}
m<\overline{\phi}\phi> \;\longrightarrow\;
<\overline{\psi}\psi> \frac{\delta_{ij}}{4}
\end{equation}
necessary on dimensional grounds. So that the above series can be
rewritten:
\begin{equation} \label{ferm8}
\begin{array}{rcl}
&&<0|:\overline{\psi}_{i\alpha}(x)\psi_{j\beta}(y):|0> \nonumber\\
&&
=\;\frac{\delta_{\alpha\beta}\delta_{ij}}{4N_c}<\overline{\psi}\psi>
      \frac{i\not\partial+m}{m}
\sum_{n=0}^\infty (-1)^n \frac{1}{n!(n+1)!}
                  \left(\frac{m^2\xi^2}{4}\right)^n \nonumber\\
&&
=\;\frac{\delta_{\alpha\beta}\delta_{ij}}{2N_c}<\overline{\psi}\psi>
      \frac{i\not\partial+m}{m^2\,|\xi|}\; J_1\left(m|\xi|\right)
\end{array}
\end{equation}

Again the commonly used formula can be recovered as lowest order term in
the mass expansion looking at eq.\,(\ref{ferm5})
\begin{eqnarray} \label{ferm9}
<0|:\overline{\psi}_{i\alpha}(x)\psi_{j\beta}(y):|0>
\;=\; \frac{\delta_{\alpha\beta}\delta_{ij}}{4N_c}<\overline{\psi}\psi>
\quad ,
\end{eqnarray}
which remains correct for vanishing mass. In the case of
non-vanishing mass the higher-order contributions of
eq.\,(\ref{ferm8}) should be taken into account.
Belyaev and Ioffe tried to explain the mass
splitting of strange and nonstrange baryons using
QCD sum rules \cite{belyaev}. It would be interesting to calculate
the importance of these mass corrections to their results.


\section{IR-divergencies and higher-order mass terms}

The relevant equation in QCD sum rules calculations is a dispersion
relation with one subtraction (see eq.\,(\ref{kap1n2})),
which gives a relation between the polarization function on the
quark level --- expanded with the use of the OPE (eq.\,(\ref{kap1n1})) ---
and the hadronic spectrum. The sum rule is obtained by taking the
Borel transformation of eq.\,(\ref{kap1n2}), which has the property
of isolating the lowest state of the hadronic spectrum (the
calculation of moments is an alternative method of isolating
the lowest state \cite{reinders}). The further the second state
in the spectrum is separated from the lowest state, the better the
isolation works.
A typical formula giving an explicit value for the mass of
the lowest state is
\begin{equation} \label{ir1}
\frac{ \int_0^{s_0} ds\;s\,  Im\Pi_{OPE}(s)\,e^{-st} }
     { \int_0^{s_0} ds\;\;\; Im\Pi_{OPE}(s)\,e^{-st} }
\;\approx\; M^2
\end{equation}
where $t$ is the Borel transformation parameter.

The main point in this formula is the sole appearance of the imaginary
part of $\Pi_{OPE}$, so that one needs only to
calculate the terms being relevant for the imaginary part. The typical
structure of a dimensional-regularized Feynman graph $G(q^2)$
($\eps=4-d$) is:
\begin{equation} \label{ir2}
G(q^2) \sim \left\{ \frac{2}{\eps} + const.
                        + ln\left(-\frac{4\pi\mu^2}{q^2}\right)
                \right\} P(q)
\end{equation}
where $P(q)$ is a purely polynomial expression and $\mu$ is the
renormalization parameter.
Remembering the analytic continuation to $q^2>0$
the whole imaginary part of such a graph is hidden in the logarithm:
\begin{equation} \label{ir3}
ln\left(-\frac{4\pi\mu^2}{q^2}\right)
\;=\; ln\left|\frac{4\pi\mu^2}{q^2}\right| + i\pi\Theta(q^2)
\end{equation}

\begin{figure}[htb]
\center{\hspace*{0mm} \epsfxsize=9.3cm \epsfysize4cm
        \epsfbox{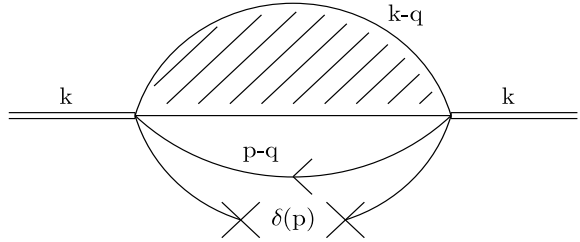}}
\caption{\label{scalarkonden} Feynman diagram with a scalar propagator, a
condensate of scalar fields and an arbitrary additional part}
\end{figure}

In the previous section a complete expression for nonlocal field
products to all orders in the mass was derived and the notion of a
{\it nonperturbative propagator} was introduced. In the following we
show that in general it is justified to restrict the calculation of
a Feynman diagram containing a condensate to a finite number of
higher-order corrections in the mass. To illustrate the argument we treat
the case of one condensate of scalar fields, which represents a large
class of graphs of the same structure. The corresponding graph is
shown in figure \ref{scalarkonden}.
The broken line represents the condensate, the full line is a
scalar propagator and the hatched area represents an arbitrary
additional structure. This could be a massless fermion loop,
for example, which
would occur in the sum rule of the $2$ quark -- $2$ squark system
introduced in the second section and which corresponds to
eq.\,(\ref{term3}). The corresponding algebraic expression
$G(k^2)$ including the mass corrections to all orders reads
\begin{eqnarray} \label{ir4}
G(k^2)
&\sim& <\overline{\phi}\phi>
    \int \frac{d^4q}{(2\pi)^4}\; f\left((k-q)^2\right)    \nonumber\\
&&
\underbrace{
    \int \frac{d^4p}{(2\pi)^4} \;
    \Delta(p-q) \,
    \sum_{n=0}^\infty \frac{1}{n!(n+1)!}
    \left(\frac{m^2\Box_p}{4}\right)^n \delta^4(p)
}_{I_p:=}
\end{eqnarray}
where the Fourier transformation of eq.\,(\ref{cond13}) has been used
and the symbol
$\Box_p = \left(\frac{\partial}{\partial p_0},-\nabla_p\right)\skal
          \left(\frac{\partial}{\partial p_0},-\nabla_p\right)$
was introduced. The function $f$ represents the hatched area in figure
\ref{scalarkonden} and $\Delta(p-q)$ is the scalar propagator.

To evaluate the integral $I_p$ it is useful to reflect once more
on the significance of the notion of the {\it nonperturbative propagator}
$\Delta_{NP}$ given by the series (\ref{cond13}). The nonperturbative
propagator satisfies the free Klein-Gordon equation so that the
relation
\begin{equation} \label{ir5}
\int d^4p\,g(p^2,p,\ldots)\,\Delta_{NP}(p)
\;=\;
\int d^4p\,g(m^2,p,\ldots)\,\Delta_{NP}(p)
\end{equation}
holds for an arbitrary function $g(p^2,p,\ldots)$ in lowest order of
perturbation theory. Note that the calculations already before
were restricted to lowest
order of perturbation theory, so that this is no additional
restriction. By applying the replacement rule (\ref{ir5}) to
the scalar propagator in eq.\,(\ref{ir4}), it can
be reduced to
$\Delta(p-q)\;=\;\left[q^2-2pq+i\epsilon\right]^{-1}$. The
$n$-th order differentiation of the distribution $\delta(p)$ is
carried out by $n$ differentiations of the simplified propagator and
after some algebra eq.\,(\ref{ir4}) reads:
\begin{equation} \label{ir6}
G(k^2)
\;\sim\; <\overline{\phi}\phi>
    \int \frac{d^4q}{(2\pi)^4}\; f\left((k-q)^2\right)
    \sum_{n=0}^\infty \frac{(2n)!}{n!(n+1)!}\,
    \frac{m^{2n}}{\left(q^2+i\epsilon\right)^{n+1}}
\end{equation}
A similar argument can be repeated for any other function $g$
instead of $g(p)=\Delta(p-q)$. In particular a fermion propagator would
lead to a similar result.

The infrared and ultraviolet behaviour of eq.\,(\ref{ir6}) is
determined by the lowest and highest power ($m_l$ and $m_h$)
of $q$ in the function $f\left((k-q)^2\right)$. So $G(k^2)$ is
UV-divergent if $n\,\le\,\frac{m_h + 2}{2}$ is fulfilled, while it is
IR-divergent if $n\,\ge\,\frac{m_l + 2}{2}$ is fulfilled. One
important conclusion can be drawn: only a finite number of terms containing
UV-divergencies exists. There is a problem region
$\frac{m_l+1}{2} \,\le\; n \;\le\,\frac{m_h+1}{2}$ with both UV-
and IR-divergencies occuring, and finally an infinite number of
terms with possible IR-divergencies exists.

To illustrate the above rules, the $2$ quark -- $2$ squark system
introduced in the second section is treated explicitly. In this case the
function $f\left((k-q)^2\right)$ in eq.\,(\ref{ir6}) becomes the
regularized massless fermion loop diagram, so that
eq.\,(\ref{ir6}) now reads:
\begin{eqnarray} \label{ir7}
G(k^2)
&=& -g^2\frac{8 N_c \pi^2}{3(2\pi)^4} <\overline{\phi}\phi>
    \sum_{n=0}^\infty \frac{(2n)!}{n!(n+1)!}\,(m^2)^n \nonumber\\
&& \times\,
   \underbrace{
    i\int \frac{d^4q}{(2\pi)^4}\,
    ln\left(\frac{(k-q)^2}{-4\pi\mu^2}\right)
    \frac{(k-q)^2}{(q^2+i\epsilon)^{n+1}}
    }_{I:=}
\end{eqnarray}
where all proportionality factors have been reintroduced.
The constant term in eq.\,(\ref{ir2}) is omitted, because
it does not contribute to the imaginary part of $G(k^2)$. The above rule
to analyze the divergencies occuring in this expression leads to the
following expectation: $f((k-q)^2)\,\sim\,(k-q)^2$ has the lowest and
highest power $m_l=0$ and $m_h=2$, so that
UV-divergencies occur for $n\le 2$, IR-divergencies occur for $n\ge 1$,
and double divergencies occur for $n\in \{1,2\}$. This is exactly
the result of the straightforward regularization of the last
integral in the expression (\ref{ir7}), which can be verified by counting
the number of divergent $\Gamma$-functions in the limit
$\eps\rightarrow 0$:
\begin{eqnarray} \label{ir8}
G(k^2)
&=& g^2\frac{8 N_c \pi^4}{3(2\pi)^8} <\overline{\phi}\phi>
    \sum_{n=0}^\infty \frac{(2n)!}{n!(n+1)!}\,(m^2)^n\,(k^2)^{2-n}
\nonumber\\
&&  \times\,
    \left(-\frac{4\pi\mu^2}{k^2}\right)^{\frac{\eps}{2}}
    \frac{\Gamma\left(1-n-\frac{\eps}{2}\right)\,
          \Gamma\left(3-\frac{\eps}{2}\right)\,
          \Gamma\left(-2+\frac{\eps}{2}+n\right)}
         {\Gamma(n+1)\,\Gamma(4-n-\eps)}
\end{eqnarray}
The double divergent terms for $n=1,2$ lead to a
divergent imaginary part, for the integral $I$ in
eq.\,(\ref{ir7}) is IR- and UV-divergent at the same time,
which leads to meaningless results for the sum rule. A more detailed
examination of this integral is unavoidable. Decomposing $I$ in three
parts, in which the integrand is proportional to $k^2$, $k\cdot q$
and $q^2$ respectively, one can easily verify that the only part
having IR- and UV- divergencies is the $k^2$--term.
It turns out that the IR-divergency in this part
can be regularized by introducing a small
IR-cutoff parameter $a^2$ into the scalar propagator (or equivalently
using a lower integration limit), reading:
\begin{eqnarray} \label{ir10}
I_{k^2}\;=\;
    i\int \frac{d^4q}{(2\pi)^4}\,
    ln\left(\frac{(k-q)^2}{-4\pi\mu^2}\right)
    \frac{k^2}{(q^2-a^2+i\epsilon)^{n+1}}
\end{eqnarray}
This integral has no IR-divergent part any more, so that it is
possible to regularize the remaining UV-divergency in the dimensional
regularization scheme. The result to order $\cal{O}(\eps)$ (where
$\eps=4-d$) depends on the cutoff parameter $a^2$:
\begin{eqnarray} \label{ir11}
I_{k^2}
&=& - \frac{\pi^2 k^2}{(2\pi)^4}
\left\{ \frac{4}{\eps^2} + \frac{2}{\eps}(1-\gamma) + const.
       + \frac{\eps}{2}\gamma^2
         ln\left(\frac{-k^2}{16\pi\mu^2}\right)
\right. \nonumber\\
&&\left. \qquad \quad
       + \left(\frac{1}{2}+\frac{\eps}{2}\gamma\right)
         ln^2\left(\frac{-k^2}{16\pi\mu^2}\right)
       - \frac{\eps}{2}
         ln^2\left(\frac{a^2}{16\pi\mu^2}\right)
\right\}
\end{eqnarray}
The essential point of this result is the exact cancellation of all
terms to the order $\cal{O}({\bf 1})$ or lower which depend on the
cutoff parameter $a^2$. The imaginary
part relevant for the sum rule in this way
remains cutoff parameter independent. The whole imaginary part of
eq.\,(\ref{ir7}) becomes finite:
\begin{equation} \label{ir9}
Im\{G(k^2)\}
\;=\; \frac{8 N_c \pi^5 g^2 m^2 k^2}{3 (2\pi)^8} <\overline{\phi}\phi>
       \left\{\frac{3}{2}
             + ln\left|-\frac{16\pi\mu^2}{k^2}\right|
       \right\}
       \,\Theta(k^2)
\end{equation}
This confirms the expectation that the IR divergencies are unphysical.

Independently of the choice of the function $f$ in eq.\,(\ref{ir6})
it must be possible to get rid of all IR-divergent terms.
This becomes clear when we look back to the very beginning. The OPE is the
basic formula of QCD sum rules and its idea is to separate the long-
and short-distance effects of nonperturbative QCD. The short-distance
effects are completely embodied in the Wilson coefficients just
calculated. So any remaining IR-divergency occuring while
calculating Wilson coefficients proves an admixture of
long-distance effects, which indicates that the OPE did not really
separate the two scales of QCD. Novikov et al. argued \cite{novikov}
that for the consistent use of OPE one has to divide the integration
domain (while calculating the Wilson coefficients) into two parts by
introducing the normalization point. In this way one can get rid of
all IR-divergencies, for only the {\it high} momentum part remains
relevant for the Wilson coefficients, like it was demonstrated in the
above example. The remaining question is wether or not the resulting
coefficients depend on the normalization point. A meaningful OPE or a
meaningful sum rule can be guaranteed only if the result is not or
weakly dependent on the normalization point.

Supposing that the Wilson coefficients are independent of the
normalization point separating the scales, one has
the possibility to restrict the
calculation of the coefficients to the terms containing an
UV-divergent part. And the above rule $n\le \frac{m_h +2}{2}$ shows
that this is a finite number of terms. A confrontation with
IR-divergent terms is inevitable only if we calculate the double divergent
terms
$\frac{m_l+1}{2} \,\le\; n \;\le\,\frac{m_h+1}{2}$ where the UV and
IR divergent parts have to be separated carefully (for example by
introducing an IR-cutoff, like it was done in the last example).

Summarizing, there is only a {\it finite} number of mass corrections in
eq.\,(\ref{cond13}) to be included in the calculation of the Wilson
coefficients, for only a finite number of terms is UV-divergent and
the IR-divergent terms cannot contribute to the Wilson coefficient
if --- and only if --- the OPE is meaningful. But this last
question leads us to the main problem of enlarging the concept of
QCD sum rules to finite mass particles. A particle of several $100MeV$
like the strange quark introduces a new scale in the theory so that the
separation of long- and short-distance effects in an OPE becomes much
more difficult than in the massless case and remains a delicate task.

\section{Conclusions}

The application of QCD sum rules to bound states composed of
particles with finite mass leads to a correction of the vacuum
expectation value of the nonlocal normal-ordered field product, which
does not remain simply a vacuum condensate (eq.\,(\ref{cond14}) for scalar
fields and eq.\,(\ref{ferm5}) for fermions) like it does in the massless
case. We derived higher-order corrections and the results
are given in eq.\,(\ref{cond13}) for scalars and
eq.\,(\ref{ferm4}) for fermions to all orders in the mass .
The importance of higher-order terms
grows in the same measure as
the mass ratio of the constituent particles and the bound
state mass.
These investigations are of great importance for the application of
QCD sum rules to mesons and baryons with strangeness.

Analysing the structure of the Wilson coefficients corresponding to
the higher order mass corrections, we found that in general a
restriction to a finite number of corrections leads not to any
neglection of mass effects, for only a finite
number of corrections is UV divergent. On the other hand, the number
of IR divergent terms is unknown and unlimited. The problem of IR
divergencies may be circumvented by introduction of an IR-cutoff
provided that a seperation of short and long distance effects in the
operator product expansion is assured to a good degree of accuracy
and provided that the final expressions does not depend strongly on the
cutoff.

We wish to thank Dr.\,Lech Mankiewicz for very helpful discussions.
This work was supported by DFG (G.\,Hess program).

\end{document}